# Acceleration of a high-charge electron bunch to 10 GeV in a 10 cm nanoparticle-assisted wakefield accelerator

Constantin Aniculaesei[1,*,#], Thanh Ha[1,#], Samuel Yoffe[5], Lance Labun[1,2], Stephen Milton[2], Edward McCary[1], Michael M Spinks[1], Hernan J. Quevedo[1], Ou Z. Labun[1,2], Ritwik Sain[1], Andrea Hannasch[1], Rafal Zgadzaj[1], Isabella Pagano[1,3], Jose A. Franco-Altamirano[1], Martin L. Ringuette[1], Erhart Gaul[1], Scott V. Luedtke[4], Ganesh Tiwari[7], Bernhard Ersfeld[5], Enrico Brunetti[5], Hartmut Ruhl[6], Todd Ditmire[1], Sandra Bruce[1], Michael E. Donovan[2], Michael C. Downer[1], Dino A. Jaroszynski[5], Bjorn Manuel Hegelich[1,2,%]

[1.] University of Texas at Austin, Austin, Texas, 78712, USA

[2.] Tau Systems Inc., Austin, Texas, 78701, USA

[3.] Lawrence Livermore National Laboratory, Livermore, California, 94550, USA

[4.] Los Alamos National Laboratory, New Mexico, 87545, USA

[5.] SUPA Department of Physics, University of Strathclyde, Glasgow, Scotland, G4 0NG, UK

[6.] Ludwig-Maximilians-Universität, Munich, Germany

[7.] Brookhaven National Laboratory, Upton, New York,11973, USA

[*] corresponding author email address: 10gevlab@gmail.com

[%] email address: hegelich@physics.utexas.edu

[#] These authors contributed equally to the work

## Abstract

An intense laser pulse focused into a plasma excites nonlinear plasma waves. Under proper conditions, electrons from the background plasma are trapped in the plasma wave and accelerated to ultra-relativistic velocities. This scheme is called a laser wakefield accelerator. We present recent results from a laser wakefield acceleration experiment that uses a petawatt-class laser to excite the wakefields and nanoparticles to assist the injection of electrons into the accelerating phase of the wakefields. We find that a 10-centimeter-long, nanoparticle-assisted laser wakefield accelerator can generate 340 pC, 10±1.86 GeV electron bunches with 3.4 GeV RMS convolved energy spread and 0.9 mrad RMS divergence. It can also produce bunches with lower energy in the 4-6 GeV range.

## I. Introduction

Since Tajima and Dawson's proposal in 1979 [1], the Laser Wakefield Acceleration (LWFA) concept has held the promise of shrinking km-scale conventional accelerators and radiation sources down to room-size machines. The LWFA utilizes a focused short-pulse laser passing through a low-density gas. The laser ionizes the gas, and the laser ponderomotive force, which is proportional to the laser intensity gradient, diverts the plasma electrons around the highest intensity regions of the laser pulse producing nonlinear plasma waves (NPW) [2]. The plasma electrons form a dense sheath around the quasi-stationary ions and create in the plasma what has been called the "bubble" [3] or "blowout" [4] regime. The trajectories of the sheath electrons collapse radially back onto themselves at the back of the bubble. The large space-charge force then pushes some of these electrons forward into the NPW, where they can be accelerated to relativistic velocities by the strong acceleration gradients. The LWFA acceleration gradients are roughly three orders of magnitude higher than those obtained by conventional radio-frequency accelerator technology, which explains the great interest in them.

Experiments exploiting the LWFA concept, as proposed initially by Tajima and Dawson, began in the late 1990s [5], [6] when chirped pulse amplification [7] using Ti:Sapphire lasers [8] produced intense TW-class femtosecond laser pulses [9]. The first quasi-monoenergetic electron bunches from an LWFA [10], [11], produced in 2004, paved the way for the generation of high-quality [12], [13], and high energy [14]–[16] electron bunches from these compact lasers.

Due to the nonlinearity of the LWFA process, the injection of electrons (number and position) into the wakefield depends very strongly on the laser and gas conditions before interaction. In an LWFA, shot-to-shot beam stability is a serious challenge. Having stable electron accelerator performance is a key requirement in developing viable applications. Small variations of laser and gas conditions lead to shot-to-shot fluctuations of the accelerated electron beam properties. Various schemes have been developed to address and control the stability of LWFA, among those worth mentioning are ionization injection [17], which increases the charge; fast down-ramp injection [18], which reduces the energy spread and controls the electron energy; and colliding laser beams [19], which control the electron beam energy. The advantages and disadvantages of each scheme will not be discussed here as it is beyond the scope of this work.

As the injection process appears to be the largest source of beam fluctuations, we experimentally explore in the present work an alternative method to inject electrons into the NPW using nanoparticles. The use of nanowires and nanoparticles has been shown theoretically [20], [21], and experimentally [22] that trigger the injection of electrons into the NPW and increase the charge density, thus providing another possible method for controlling the parameters of the accelerated electron beam. In our experiment, the nanoparticles are generated inside a gas cell through laser ablation of a metal surface and are assumed to be mixed uniformly with the helium gas fed inside the gas cell. That said, we cannot yet control when the injection happens due to the random distribution of the nanoparticles in the current experiment. The combined control over how and where the electron injection happens could be achieved using an aerodynamic lens [23], for instance. Developing and integrating such an aerodynamic lens into a gas target would require significant financial and human resources; thus, in this first instance, we focus on the usefulness of nanoparticle-assisted wakefield acceleration, which is the purpose of the experimental results presented in this work.

## II. Experimental Method

### 1. Overview

The experimental setup is shown in *Figure 1*.

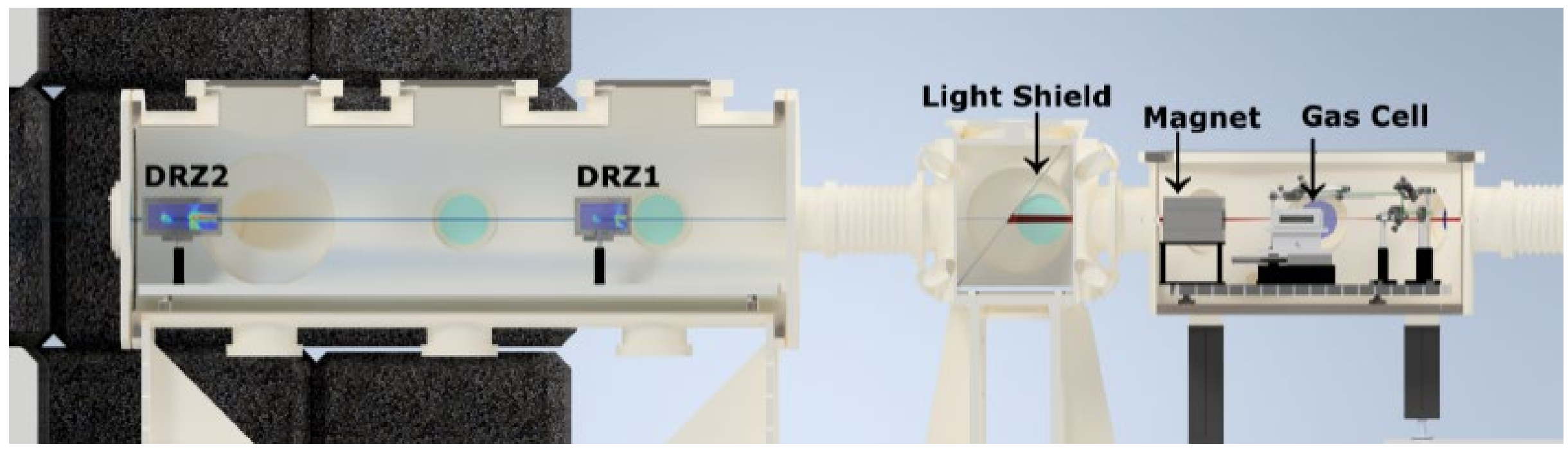

*Figure 1: The electron diagnostics setup containing the gas cell, a dipole magnet, and two scintillating screens, DRZ1 and DRZ2. The entire setup is placed in vacuum chambers. The laser and electron bunches propagate from right to left.*

An f/50 spherical mirror focuses the intense petawatt-level laser pulses (135 fs pulse duration and 130 J energy) inside a 10-cm long gas cell filled with 99.9% purity helium and doped with aluminum nanoparticles. The leading edge of the laser pulse ionises the gas, creating a plasma with an electron density of $6\pm0.5 \times10^{17}$ cm$^{-3}$; concurrently, the peak of the laser pulse excites an NPW in the bubble regime providing the conditions to accelerate electrons to high energy. The electron bunches from the LWFA are subsequently deflected by a 10-cm-long dipole magnet with a B-field of 0.79 T and detected on three scintillating screens at 1.568 m, 2.556 m, and 5.855 m, respectively, downstream of the exit pinhole of the gas cell. As detailed in the **Electron beam diagnostics** subsection, multiple screens allow cross-checking and accurate reconstruction of the electron energy spectra independent of the initial pointing of the electron beam. The beam distribution spread measured on the screens includes components from the electron beam energy spread and the electron beam divergence. This beam divergence term thus limits the energy spread resolution of the electron spectrometer as it is not deconvolved due to the lack of simultaneous measurement of the electron beam divergence in the bend plane.

The farthest screen, DRZ3 (shown in *Figure 3*), detects electrons with energies above 2 GeV, whilst the closer two screens, DRZ1 and DRZ2, detect electrons with energies above 0.4 GeV. As detailed in the following section, we used an imaging plate and cross-correlation with the light emitted by the scintillating screens for charge calibration.

## 2. The Texas Petawatt Laser

The Texas Petawatt Laser delivers 130 ± 10 J pulses on target with 45% of the total energy enclosed within $1/e^2$. The FWHM pulse duration is 135±10 fs with a central wavelength at 1057 nm. An f/50 spherical mirror focuses the laser pulse to a FWHM focal spot of ~55 μm and a peak intensity of $1.2 \times 10^{19}$ W/cm$^2$. The laser temporal contrast of the laser pulse, up to several tens of picoseconds before the main pulse peak, has been measured to be on the order of $10^{-8}$. The laser parameters are monitored before hitting the spherical mirror, and its energy, Strehl ratio, collimation, etc., are retrieved for each shot. More details on the TPW laser construction and performance can be found in the published literature [24]–[26].

## 3. Gas target and nanoparticle source

The gas target [27] 3D drawing is shown in *Figure 2*. Its design is based on the *SlitCell* design [28], modified to accommodate a removable metal plate on the bottom of the gas cell for nanoparticle generation. The gas cell has two windows, one on the side and another on the top, used for laser alignment and visualization of the interaction region. The gas target is filled with helium via a solenoid valve opening for 2 ms and delayed 27 ms before the main laser arrives. The gas density is monitored with a pressure transducer (measured standard deviation of $0.5 \times 10^{17}$ $cm^{-3}$) installed in the middle of the gas cell wall. According to fluid dynamic simulations (not shown here), the gas density profile is uniform inside the gas cell and presents down-ramps outside the pinholes.

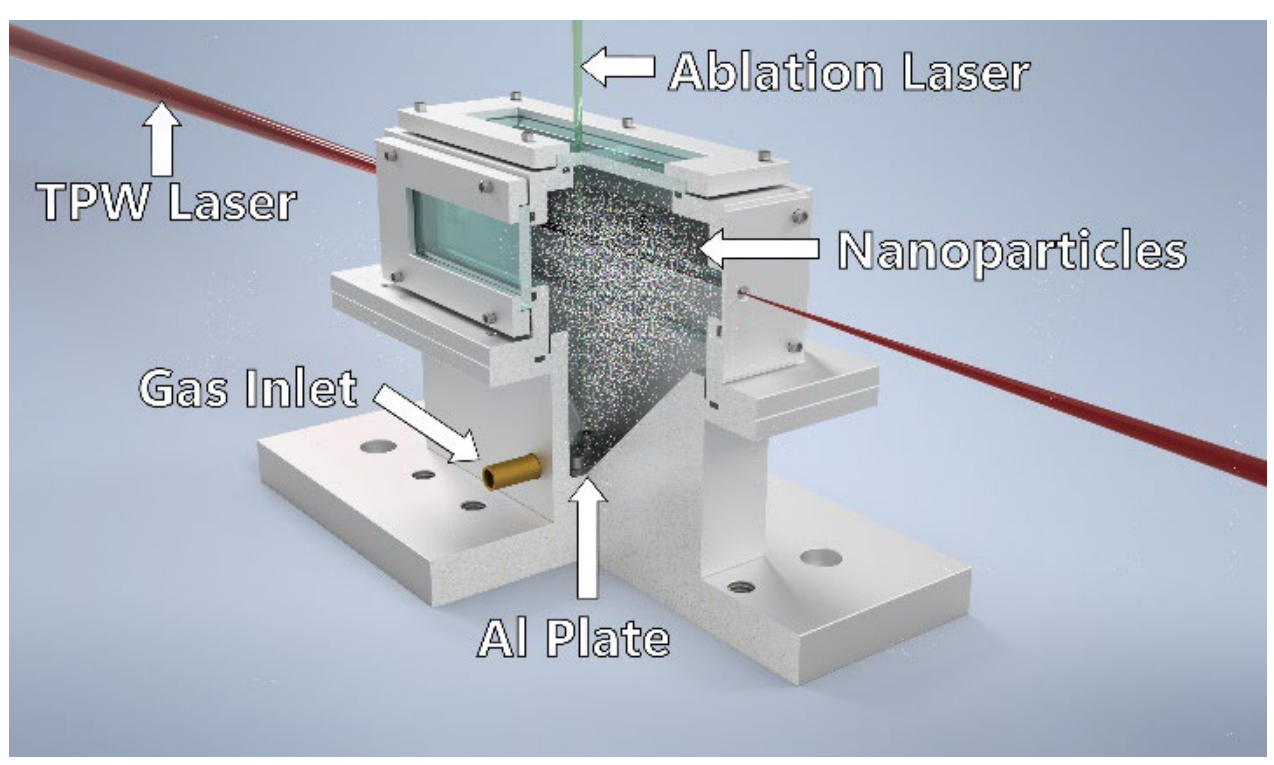

*Figure 2: Gas cell drawing. A 532 nm laser is focused through the top window onto the surface of a metal plate and generates the nanoparticles through laser ablation. The nanoparticles mix with the helium gas and fill the volume of the gas cell uniformly. The Texas Petawatt Laser enters the gas cell through a 3 mm diameter pinhole and generates electrons which exit the gas cell through another 3 mm pinhole.*

In shots with nanoparticles, an auxiliary laser pulse (532 nm wavelength, 10 ns pulse duration, and 130 mJ energy) is fired 500 microseconds prior to the main petawatt pulse onto an aluminium plate situated on the bottom of the gas cell near the gas inlet, ablating it and creating the nanoparticles [29], [30]. Theoretical [21] and experimental [31] investigations have shown that the amount of charge injected in the bubble can be controlled by changing the nanoparticle's composition, size, or density. We used an aluminium plate for the work presented here, but most metals can be used as a solid plate or deposited on a support plate. The nanoparticles mix with the helium gas to fill the gas cell uniformly. The ablation laser fluence is kept constant at 5 $J/cm^2$ for the entire experiment. We estimate using reference [30] that the mass of ablated aluminium per shot is m = 19 μg. For simplicity, we assume that the entire ablated mass is transformed into nanoparticles with a 10 nm diameter [31] uniformly distributed over the entire volume of the gas cell. In this case, $\sim 10^6$-$10^7$ nanoparticles can interact with the laser in a cylinder defined by the 50 μm laser spot diameter and the 10 cm gas cell length. While this provides a rough upper-bound estimate for the nanoparticle density, simplifying assumptions in this calculation leads us to believe that the actual nanoparticle density may be lower by a few orders of magnitude.

## 4. Electron beam diagnostics: energy and charge

We deployed a multi-screen electron spectrometer [32] for the electron beam energy characterization, as shown in *Figure 3*. This arrangement consists of a known static magnetic field and multiple scintillating screens to reconstruct the electron trajectories. It consists of a 0.79 T peak field dipole magnet between A and B and two DRZ1 and DRZ2 scintillating screens placed at C and D, respectively (imaged by two sCMOS cameras). Two Fuji BAS-SR imaging plates placed after DRZ1 and DRZ2 screens were used for charge calibration of the scintillating screens. An imaging plate named DRZ3 placed at E in *Figure 3* detects electrons with energies higher than 2 GeV. The electron energy retrieving algorithm published by Hojbota et al. [33] is used to determine the energy spectrum of the electron beams produced in the laser wakefield experiment.

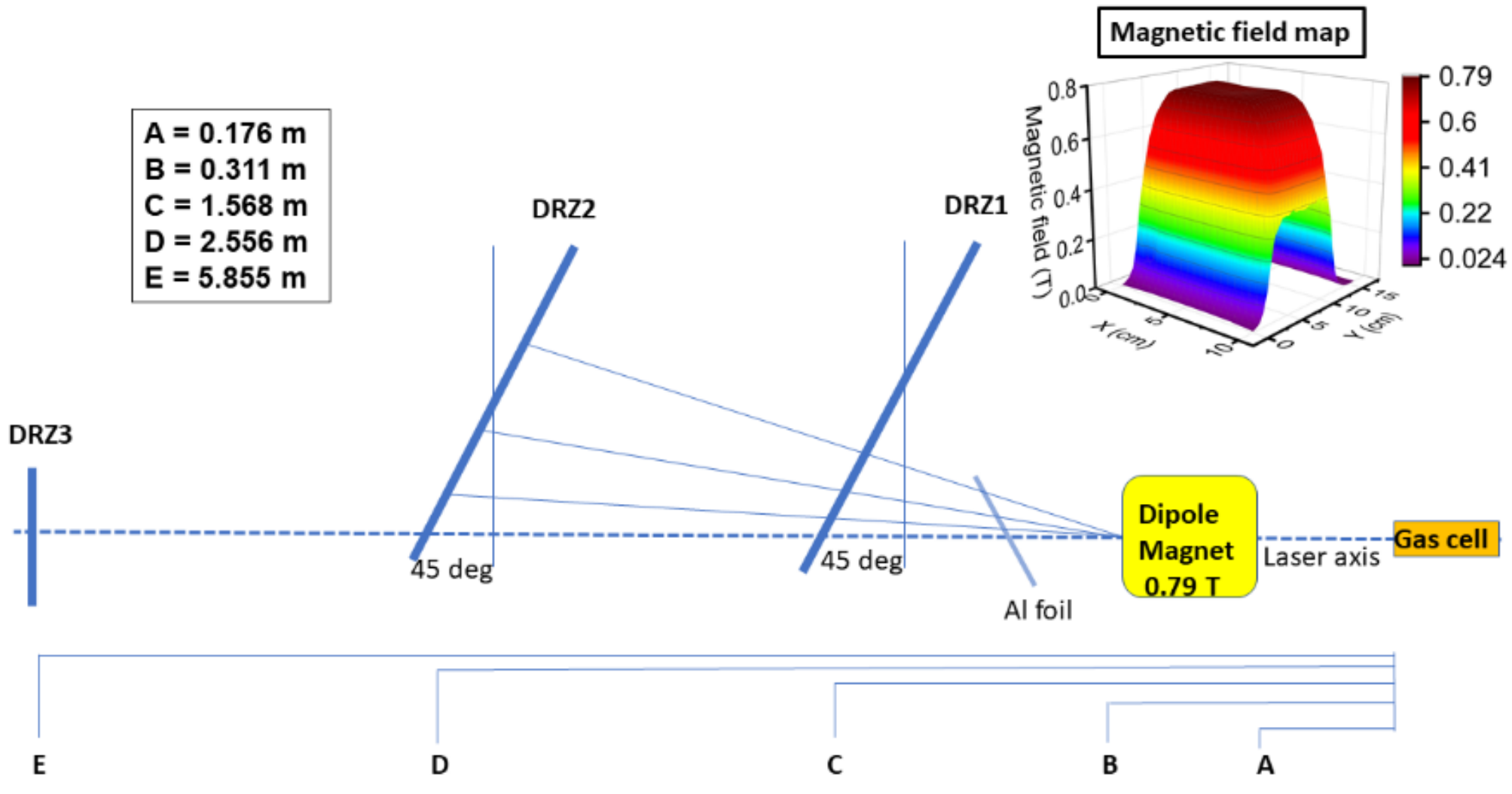

Figure 3: 2D drawing of the setup containing the gas cell and diagnostics. The inset shows the measured magnetic field map of the dipole magnet. The laser and electron bunches propagate from the right to the left.

We employed the following analysis to mitigate the uncertainty generated by the electron beam angle and offset at the entry point of the dipole magnet.

An electron traveling through a magnetic field oriented perpendicular to its motion is subjected to the Lorentz force:

$$\frac{d\mathbf{p}}{dt} = \frac{d}{dt}\left(\gamma m_e \mathbf{v}\right) = -e\mathbf{v} \times \mathbf{B}.$$

For simplicity, we will consider that the electron propagates solely in the $\hat{\mathbf{z}}$-direction initially before encountering the magnetic field, and the magnetic field is solely in the $\hat{\mathbf{y}}$-direction. Then any deflection imparted onto the electron by the magnetic field is in the $\hat{\mathbf{x}}$-direction. We also assume that any radiation loss due to acceleration in the magnetic field is negligible compared to the electron's initial energy. Thus, we can approximate the effects due to the magnetic field on the electron trajectory as

$$x(z)=\frac{v_x(z=0)}{v_z}z+\frac{e}{\gamma m_e v_z}\int_0^z dz'\int_0^{z'}dz''B_y(z'').$$

For a relativistic electron, the propagation speed is nearly the speed of light $v_z \approx c$. With this approximation, we can substitute the electron energy $\text{ò}=\gamma m_e c^2$ and obtain the following:

$$x(z)=\beta_{0x}z+\frac{ec}{\text{ò}}\int_0^z dz'\int_0^{z'}dz''B_y(z'').$$

The trajectory of the electron leaving the magnetic field can be determined by measurement of the electron positions on two or more scintillating screens positioned after the magnet. The electron deflection is known, while the initial trajectory pointing is unknown. We can recover the energy of the electron by using the measured trajectory and propagating the electron backward through the detector system instead of forward propagating the electron from the source into the detector. In the backward propagation, this trajectory determines $\beta_{0x}$ in the equation:

$$\beta_{0x}=-\frac{x_2-x_1}{z_2-z_1}=-\frac{\Delta x}{\Delta z}.$$

The energy of the electron producing the trajectory can be determined as

$$\text{ò}=\frac{ec}{-x_2+\frac{\Delta x}{\Delta z}z_2}\int_0^z dz'\int_0^{z'}dz''B_y(z'').$$

One must also consider the errors in the measurement process. Consider some variation in the measured deflection $\delta x_i$. The impact on the energy can be determined as

$$\delta \grave{o}_x = \left( \sum_i \left( \frac{\partial \grave{o}}{\partial x_i} \right)^2 \delta x_i^2 \right)^{1/2} = \grave{o} \left( \left( \frac{-z_2 \delta x_1}{x_2 z_1 - x_1 z_2} \right)^2 + \left( \frac{z_1 \delta x_2}{x_2 z_1 - x_1 z_2} \right)^2 \right)^{1/2},$$

where the $i^{th}$ index corresponds to the $i^{th}$ scintillating screen (See *Figure 4*).

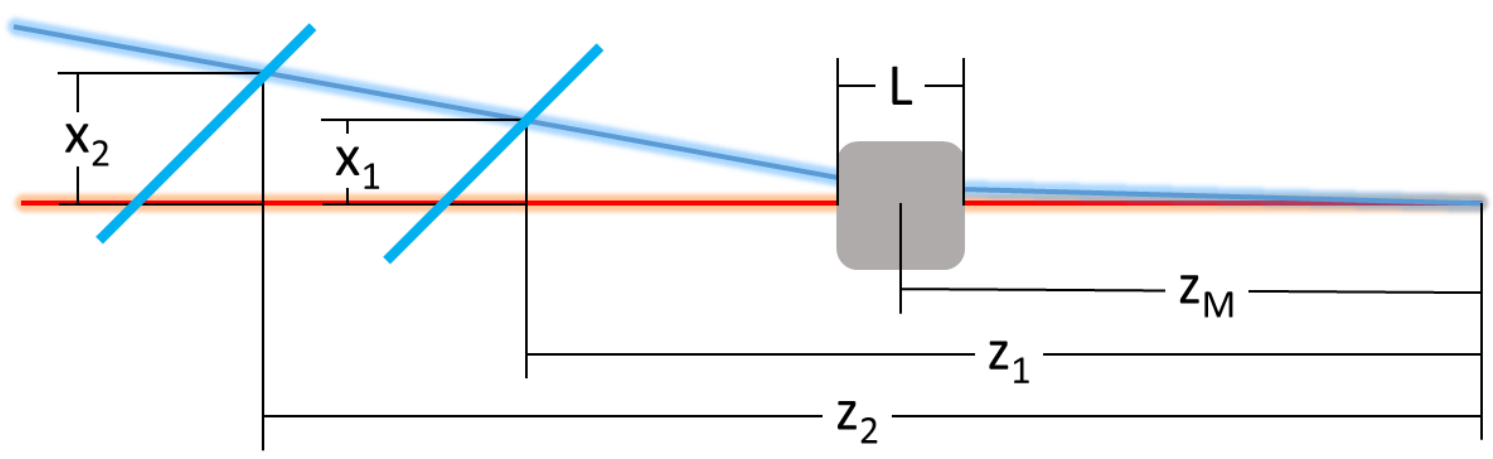

Figure 4: Simplified setup used to calculate the error in centroid electron energy. The laser and electron bunches propagate from the right to the left.

For our design, $z_1$ = 1.702 m and $z_2$ = 2.686 m, measured from the end of the gas cell to the intersection point on the scintillating screen. We expect the variation in measurement due to optical system resolution to be of the order of $\delta x_1$ = 38 μm and $\delta x_2$ = 60 μm. For ~6 GeV electron energy, we expect $x_1$ = 5.700 mm and $x_2$ = 9.765 mm with the magnet used in the work presented here. This results in $\delta \grave{o}_x / \grave{o} = 0.111$.

Similarly, we can estimate the variation in the energy measurement due to variations in screen and magnet spacing. For simplicity, we will assume the true magnetic field can be approximated by a fixed field of strength $B_0$ over some length L ± δL. Then the energy of the electron is:

$$\grave{o} = \frac{ecB_0}{-x_2 + \frac{\Delta x}{\Delta z} z_2} \left( \left( z_2 - z_M + \frac{L}{2} \right)^2 - \left( z_2 - z_M - \frac{L}{2} \right)^2 \right)$$

Where $z_M \pm \delta z_M$ is the location of the center of the magnetic field. Note that the variation in the effective field strength corresponds to a minor variation in energy,

$$\frac{\delta \epsilon_B}{\epsilon} = \frac{\delta B}{B_0} = 0.0013$$

for a field strength of $B_0 = 0.760$T and $\delta B = 0.001$ T (instrument measurement error). If we look at the variation in the propagation axis z, the variation in the energy can be quantified as

$$\frac{\delta \varepsilon_z}{\varepsilon} = \left( \left( \frac{z_2 (x_1 - x_2) \delta z_1}{(z_2 - z_1)(x_2 z_1 - x_1 z_2)} \right)^2 + \frac{4(z_2 - z_M)^2 \delta L^2 + 4L^2 \delta z_M^2}{\left( (z_2 - z_M + L/2)^2 - (z_2 - z_M - L/2)^2 \right)^2} \right.$$

$$\left. + \left( \frac{z_1 (x_2 - x_1)}{(z_2 - z_1)(x_2 z_1 - x_1 z_2)} + \frac{2L}{(z_2 - z_M + L/2)^2 - (z_2 - z_M - L/2)^2} \right)^2 \delta z_2^2 \right)^{1/2}$$

For our spectrometer, the expected energy variation due to positional location measurement error $\delta z_1 = \delta z_2 = \delta z_M = \delta L = 0.5$ mm is $\delta \epsilon_z / \epsilon = 0.009$ for 6 GeV electron energy. The total error is given by

$$\frac{\delta \epsilon}{\epsilon} = \sqrt{\left( \frac{\delta \epsilon_x}{\epsilon} \right)^2 + \left( \frac{\delta \epsilon_z}{\epsilon} \right)^2} = 0.112.$$

At the lower energy limit of our spectrometer (~ 600 MeV), we expect electron trajectories with $x_1 = 54.660$ mm, $x_2 = 93.652$ mm, and $\delta x_1 = 41$ μm, $\delta x_2 = 63$ μm. The resolution is roughly the same as for the higher energy electron trajectories since we view the scintillating screen at a near-normal incidence. This yields $\delta \epsilon_x / \epsilon = \delta \epsilon_z / \epsilon = 0.012$. For the low energies, we estimate the total error as $\delta \epsilon / \epsilon = 0.017$.

The total error energy as a function of the electron energy is shown in *Figure 5*, where the maximum plotted error is 18.6% for 10 GeV electron energy.

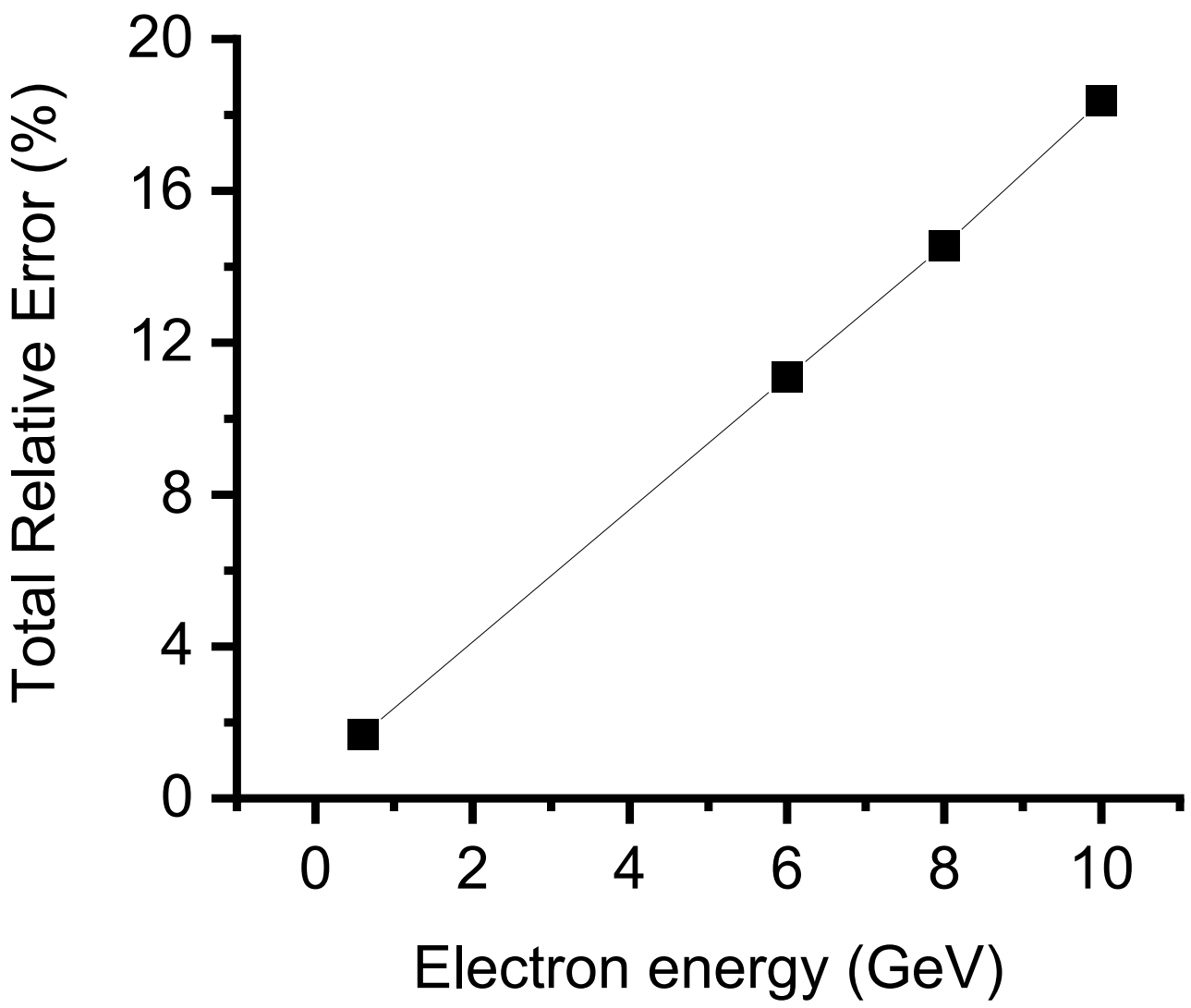

Figure 5: The electron spectrometer total relative error in energy retrieval as a function of the electron energy.

Using G4Beamlines [34], a particle simulation and tracking program, we can simulate electron trajectories through our detector using various initial parameters while including the measured dipole field map. With these simulations, we can produce the parameter space (all the positions for all possible pointings between -1 mrad and +5 mrad) for each detector scintillating screen (*Figure 6*). For each experimental shot, we obtained the positions of the bunch or bunches on each screen, and using the parameter space diagrams, we found what pointing corresponds to the same energy reading on both screens. This provides the energy calibration for plotting the shot and the spectrum. As a check and when possible, the calibration is done for multiple points to ensure that different parts of the spectrum do not have different pointing.

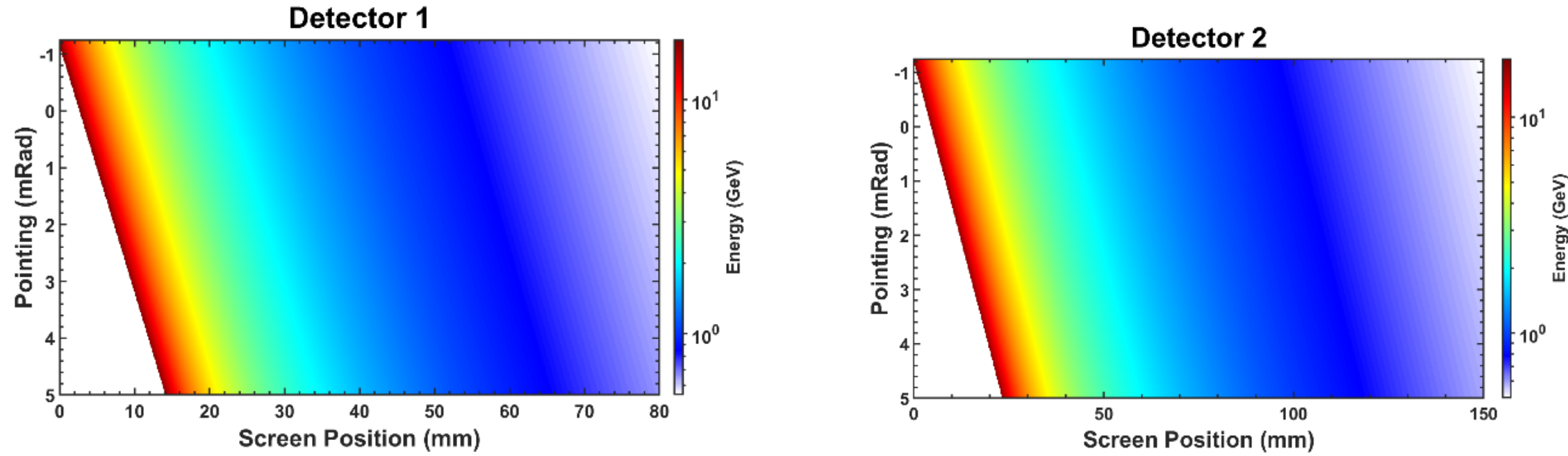

Figure 6: Parameter space for DRZ scintillating screen 1 and 2 (Detector 1 and Detector 2, respectively) for various initial electron pointing and energies. The experimental electron spectrum is matched on each screen, and the corresponding pointing and energy are retrieved for each spectrum feature or each bunch.

Following calibration protocols found in the literature [35]–[38], the imaging plates were used to cross-calibrate the electron beam charge impinging onto the scintillating screens. The charge of the electron beam is determined from the imaging plate by the formula:

$$Q = \frac{Se}{R(E)\alpha(t,N)}$$

Where S is the PSL signal from the imaging plate, e is the elementary charge, R(E) is the imaging plate response sensitivity for a given incident electron energy E, and $\alpha(t,N)$ is the convoluted attenuation of the signal due to repeatedly scanning N times and the passage of t time. Both R(E) and $\alpha(t,N)$ are empirically determined.

To obtain the charge on shots without an imaging plate, we can correlate the charge obtained from the imaging plate to the light signal captured by our optical camera equipped with a narrow band-pass filter to remove some of the background light. All shots were taken with the camera set to zero gain and a fixed exposure length. In an effort to avoid signal saturation of the CCD pixels, the aperture f-stop of the imaging lenses was set a priori best on our best estimate of the expected signal intensity. Each decrease of the aperture f-stop corresponds to a doubling of light intensity on the sensor, and inversely, each increase of the aperture f-stop corresponds to a 1/2 factor in light intensity. The f-stop for each shot was recorded, and the corresponding correction factor was applied to obtain the emission signal from the scintillating screen. As determined by shots recorded with the imaging plate signal, the scintillating screen emission signal has a linear relationship to the electron beam charge. Despite our attempts at staying out of the signal saturation regime, some electron spectra were saturated. The retrieved charge is underestimated in those cases.

## III. Results

In baseline shots without nanoparticles, we produced typical electron bunches similar to those published by Wang et al. [39] with electron energies around 2 ± 0.08 GeV and charge of a few hundred pC (see *Figure 7*).

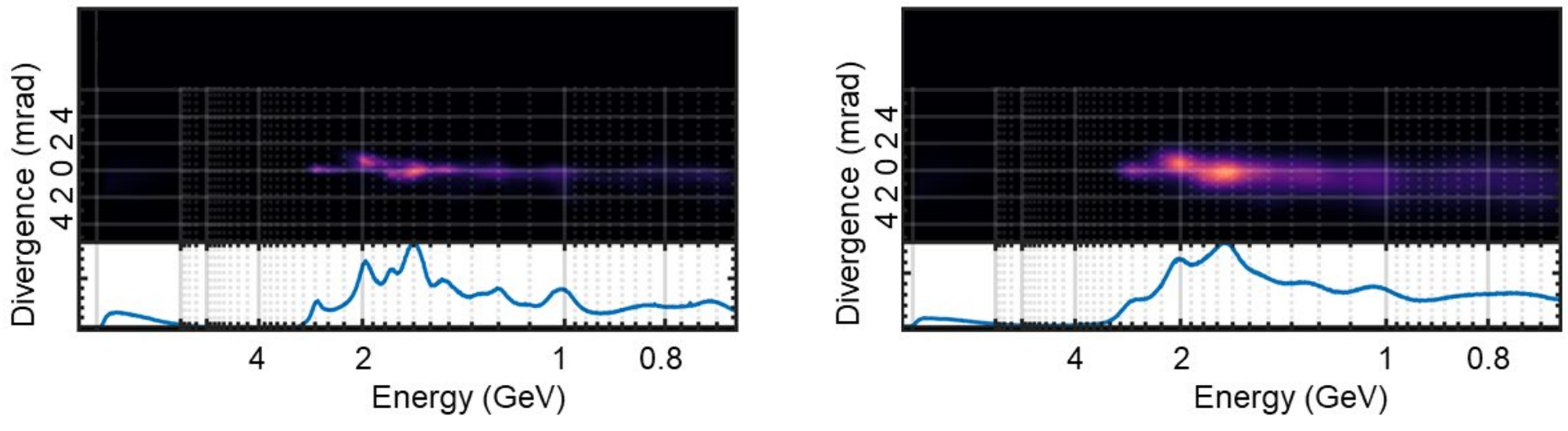

*Figure 7: Typical shot recorded without nanoparticles and shown on both successive DRZ screens. The difference in charge and divergence is due to a different response of the DRZ screen and optical system assembly.*

The low repetition rate of the TPW laser precluded systematic parameter scans, and the realities of high-power laser systems and limited beam time resulted in 26 successful shots (23 with and 3 without nanoparticles) in our experimental campaign, from which two shots with nanoparticles showed electron energies beyond 10 ± 1.86 GeV. The electron spectra that displayed the highest attained energy are shown in *Figure 8* and *Figure 9*. More electron spectra with energies beyond 2 GeV, including the already mentioned shots, with an energy lower bound cut at 2 GeV, are shown in *Figure 10* and summarized in *Table 1*.

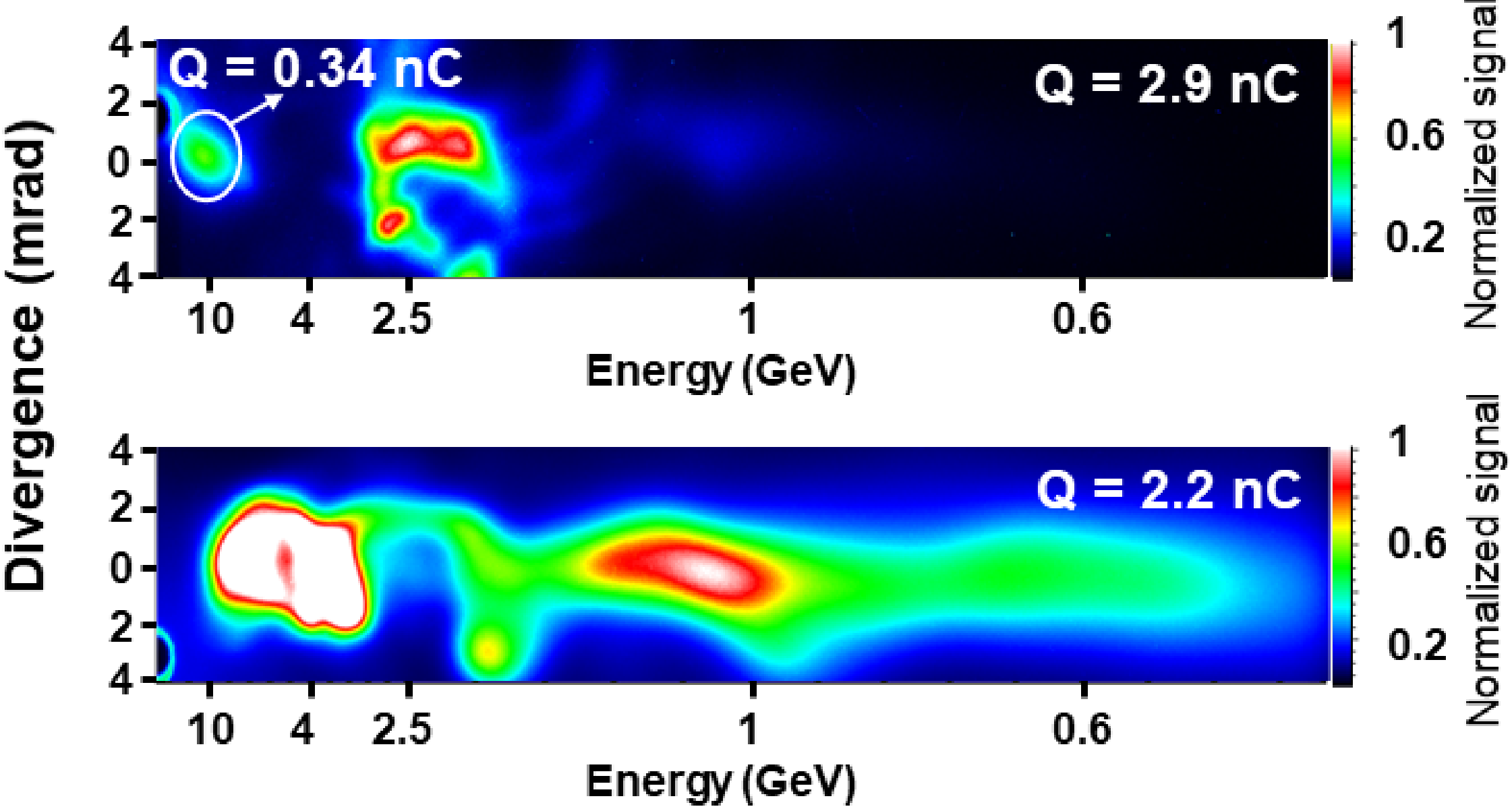

*Figure 8: Electron energy spectra of the two most energetic shots recorded by DRZ2. The energy spectra are recorded simultaneously on two consecutive screens to correct any off-axis electron beam pointing. The top spectrum shows a high energy bunch with the centroid at 10 ± 1.86 GeV, 3.4 GeV RMS energy spread, 340 pC electric charge (2.9 nC total charge), and 0.9 mrad RMS divergence. The bottom energy spectrum shows a 4.9 ± 0.39 GeV centroid electron bunch with tail energy that extends beyond 10 GeV and has a 2.2 nC total charge with 1.4 mrad RMS divergence. The energy spread from the electron beam divergence has not been deconvolved, and its value can be lower than estimated.*

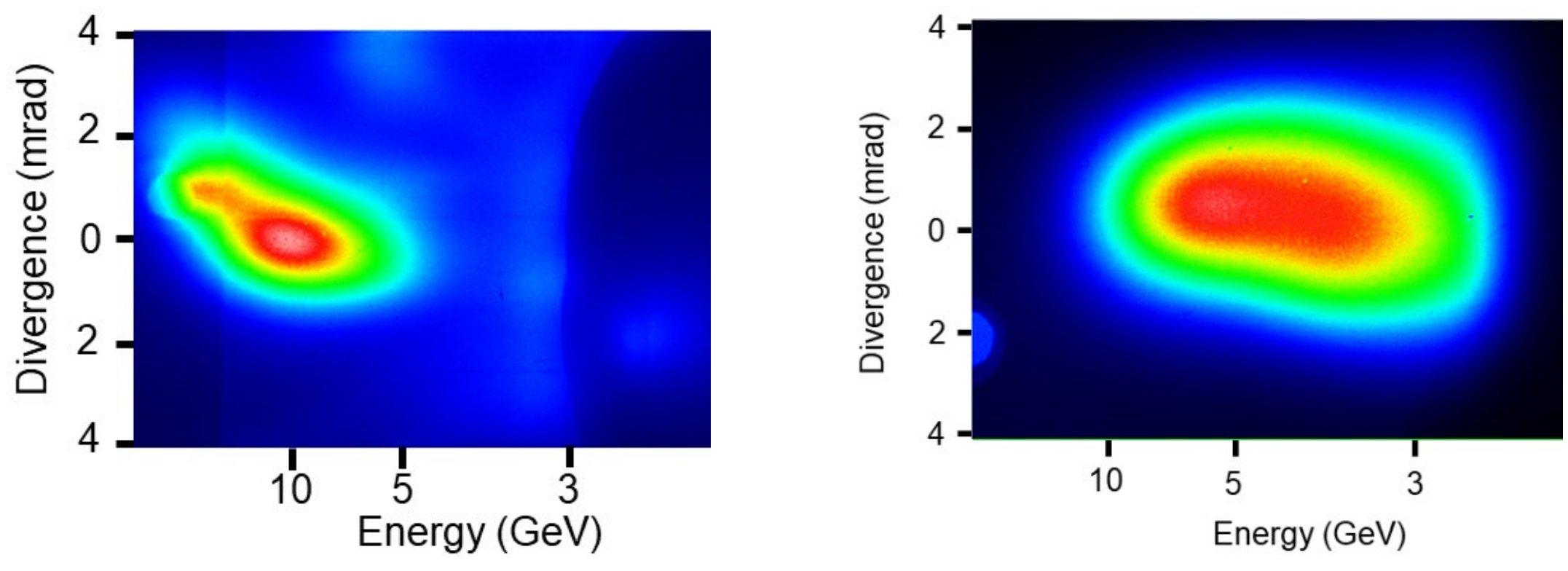

*Figure 9: Two of the most energetic electron spectra as viewed on DRZ3 (placed 5.855 m away from the exit of the gas cell) with an energy cutoff of ~2 GeV.*

As an exemplification of data analysis, let us consider shot number 3 (refer to Figure 10 and Table 1). The initial raw data exhibited two distinct electron bunches, and we accurately determined their positions on the DRZ screens. Subsequently, employing the analysis routine, we determined the unique solution which gives the pointing for each bunch while simultaneously they displayed identical energy readings on both DRZ screens. Consequently, upon comparing the pointing values for each bunch, it was revealed that both exhibited identical values of 2.20 mrad (meaning that the electron energy was lower than the real one if uncorrected). Using this corrected pointing value, we derived the electron spectrum.

Similarly, the analysis conducted on shot number 6 necessitated a pointing correction of 0.75 mrad. In contrast, the remaining shots did not require any correction.

Although not obvious in Figure 6 due to the limited optical resolution, the raw data used to generate the parameter space shows that within the experimental error, there is a unique solution that gives the same energy reading on both DRZ screens and has the same pointing.

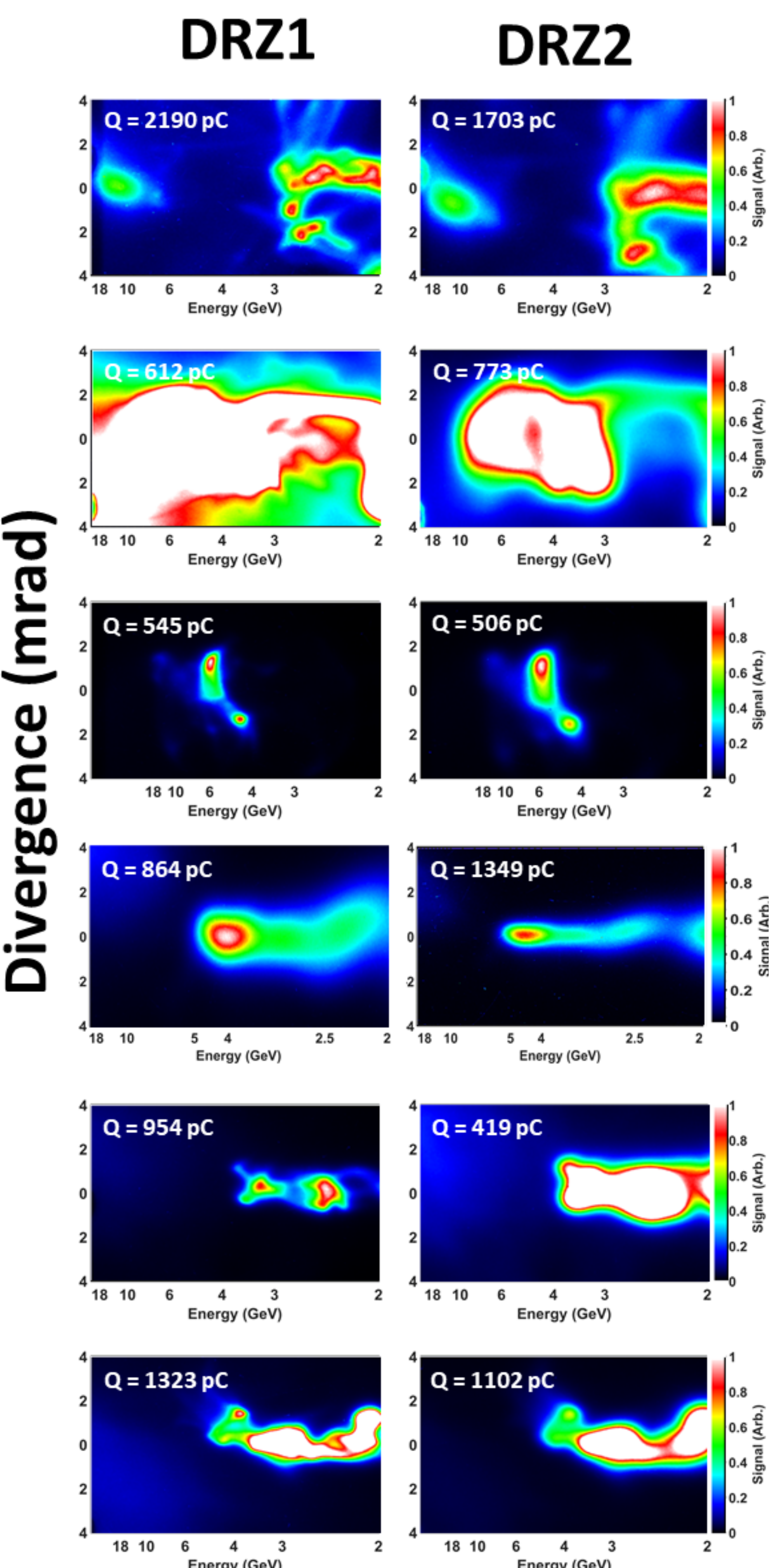

*Figure 10: Data showing the electron energy spectra with energies above 2 GeV recorded by DRZ1 (left column) and DRZ2 (right column). The DRZ1 screen is placed 1.568 m from the exit of the gas cell, and DRZ2 is placed at 2.556 m from the exit. The first two shots show the highest electron energies beyond 10 GeV.*

| Shot | Pulse duration (fs) | Laser energy (J) | Focal plane position (mm) | Strehl ratio | Electron centroid energy (GeV) | Total charge (pC) | Pointing correction (mrad) |
|---|---|---|---|---|---|---|---|
| 1 | 134 | 118 | 7.21 | 0.72 | 10±1.86 | 1703 | 0 |
| 2 | 143 | 125 | 7.05 | 0.4 | 4.9±0.42 | 773 | 0 |
| 3 | 136 | 124 | 7.05 | 0.64 | 6.2±0.68 | 506 | 2.2 |
| 4 | 147 | 97 | 4.21 | 0.58 | 4.5±0.36 | 1349 | 0 |
| 5 | 139 | 128 | 7.69 | 0.61 | 3.5±0.22 | 419 | 0 |
| 6 | 134 | 126 | 6.29 | 0.47 | 3.4±0.20 | 1102 | 0.75 |

*Table 1: The laser parameters corresponding to some of the highest electron energy shots. The electron energy is taken as the centroid of the highest energy bunch. The charge is taken from the DRZ2 with a low cutoff energy of 2 GeV.*

The position of the laser focal plane was monitored during the experiment but was not accurately controlled and presented important shot-to-shot fluctuations. We observed that the position of the focal plane inside the gas target was essential in controlling the electron energy, as observed in previous experiments with the TPW laser [40]. All electron energy spectra with peak energies beyond 3.5 GeV (all generated using nanoparticles) are obtained with the expected laser focal plane in vacuum at 7±1 mm (see *Figure 11* ) inside the gas cell relative to the entrance pinhole.

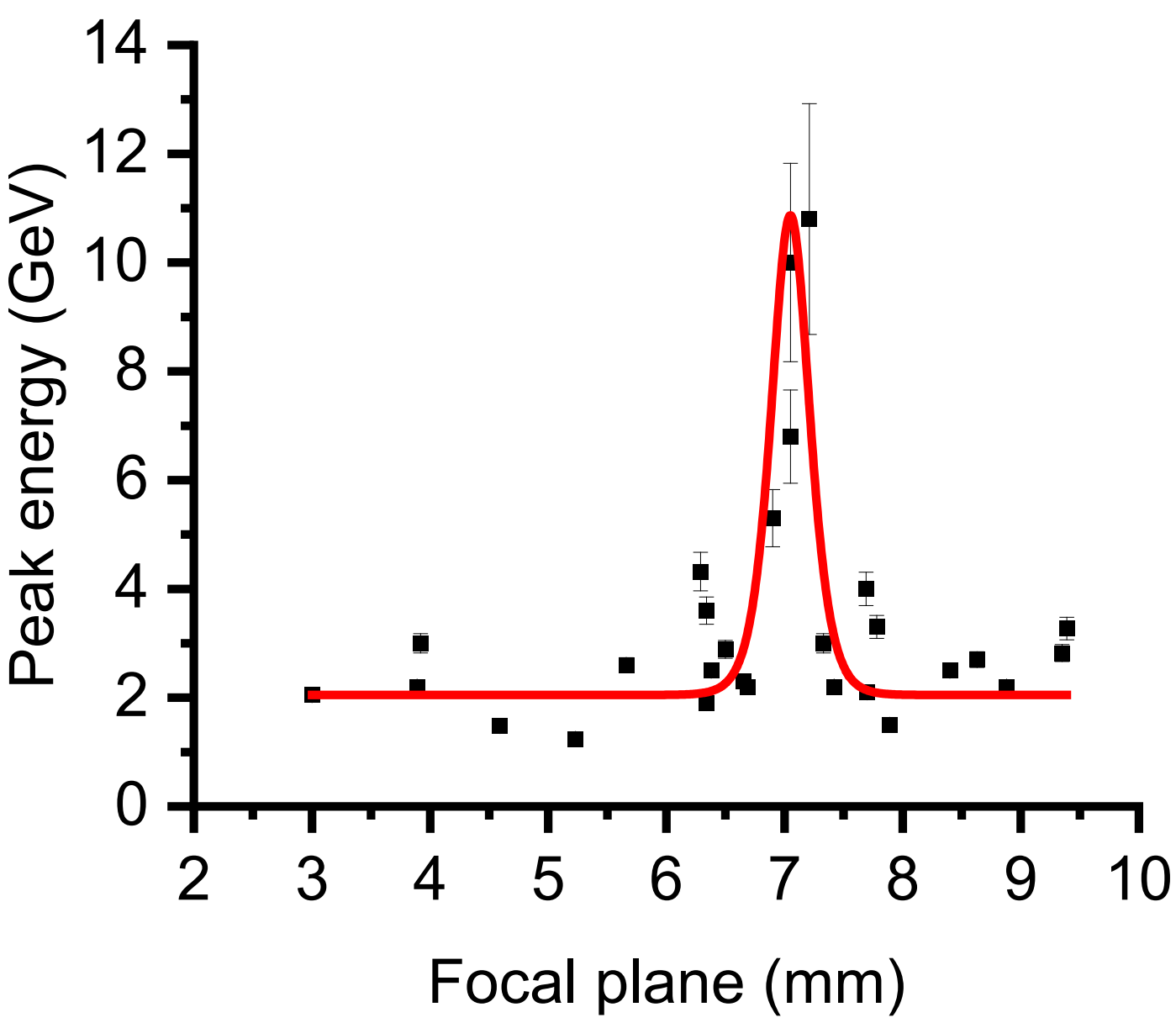

*Figure 11: Dependence of the maximum (or peak energy) electron energy on the position of the laser focal plane in the gas cell. It can be observed that all the shots with electron energies above 3.5 GeV are grouped around 7±1 mm. The red curve is drawn to guide the eye, and the entrance pinhole is at 0 mm.*

The significant reliance of the outcome of Laser Wakefield Acceleration (LWFA) on the spatial location of the laser focal plane within the gas target may be attributed to the nonlinear evolution of the laser pulse and plasma wave, contingent upon the initial plasma conditions as shown by Ciocarlan et al. [41]. The potential impact of the entrance pinhole of the gas cell on the laser beam, which has not been quantified [42], should also not be disregarded. Additional empirical and theoretical inquiries are imperative to elucidate the fundamental mechanism, which is presently being actively investigated.

## IV. Discussion and conclusion

Currently, we do not have a satisfactory model or experimental proof for generating such high electron energies. Various theoretical scenarios are now under investigation and, if relevant, will be the subject of future publication. Due to the prohibitive computational complexity and cost of performing PIC simulation in a full 3-dimensional geometry of 10 cm long plasma, with a known spatio-temporal shape laser pulse and the additional need to resolve nanoparticles (resolution less than 5 nm), our future effort will focus on characterizing the nanoparticle-assisted wakefield accelerator in terms of the output electron parameters, i.e., better statistics and by probing the wakefields using few-cycle lasers [49] and electron beams [50]. Due to the high number of shots required for this experiment (statistics and probing), access to a high repetition rate petawatt-class laser (0.01-1Hz) will be required.
In conclusion, we have shown in this nanoparticle-assisted laser wakefield acceleration experiment that we could produce electron bunches with high energies between 4-10 GeV. From the 26 recorded electron spectra under various experimental conditions, one electron spectrum shows an electron bunch with 0.34 nC of charge and centroid energy of 10±1.86 GeV, while another electron spectrum shows electron bunches with a tail extending beyond 10 GeV. Further investigations using a high repetition PW-class laser, such as the ones found at *BELLA* at Lawrence Berkeley National Laboratory, *ALEPH* at Colorado State University, CoReLS in the Republic of Korea, or *ELI-NP* in Romania, we may be able to identify the mechanisms which enable the production of 4-10 GeV electron beams and the experimental conditions required to improve their quality.

**Acknowledgements**
BM Hegelich, C Aniculaesei, T Ha, L Labun, OZ Labun, and E McCary have been supported by the Air Force Office of Scientific Research Grant No. FA9550-17-1-0264. This work was supported by the DOE, Office of Science, Fusion Energy Sciences under Contract No. DE-SC0021125:

LaserNetUS: A Proposal to Advance North America's First High Intensity Laser Research Network. The contributions of A Hannasch, R Zgazdaj, I Pagano, JA Franco, and MC Downer were supported by the U.S. Department of Energy grant DE-SC0011617. DA Jarozynski, E Brunetti, B Ersfeld, and S Yoffe would like to acknowledge support from the U.K. EPSRC (EP/J018171/1, EP/N028694/1) and the European Union's Horizon 2020 research and innovation program under grant agreements no 871124 Laserlab-Europe and EuPRAXIA (653782). Simulation results were obtained using the ARCHIE-WeSt High-Performance Computer (www.archie-west.ac.uk) based at the University of Strathclyde, and the facilities of the N8 Centre of Excellence in Computationally Intensive Research (N8 CIR) provided and funded by the N8 research partnership and EPSRC (grant number EP/T022167/1), coordinated by the Universities of Durham, Manchester, and York.

Many thanks to Rémi Lehe of Lawrence Berkeley National Laboratory for his support in deploying and optimizing the FBPIC code.

**Competing interests**

Bjorn Manuel Hegelich and Constantin Aniculaesei submitted a patent application, 17/845,223, filled on June 21, 2021, describing the device and method to generate nanoparticles in a gas cell. BMH, LL, OZL, SM, and MED are employed by the company Tau Systems Inc., which is a company that develops and sells technology based on wakefield accelerators.

**Data availability**

The data are available from the corresponding author upon reasonable request.